# The statistical distribution of money and the rate of money transference

Juan C. Ferrero[a]

*INFIQC, Departamento de Fisicoquímica, Facultad de Ciencias Químicas*

*Universidad Nacional de Córdoba, 5000 Córdoba, Argentina*

The distribution of money is analysed in connection with the Boltzmann distribution of energy in the degenerate states of molecules. Plots of the population density of income distribution for various countries are well reproduced by a Gamma function, confirming the validity of the statistical distribution at equilibrium. The equilibrium state is reached through pair wise money transference processes, independently of the shape of the initial distribution and also of the detailed nature of the money transactions between the economic agents.

Short title: statistical distribution and rate of money transference

---

[a] jferrero@mail.fcq.unc.edu.ar

# 1 Introduction

The distribution of money between the economic agents in a society and the rate at which this process occurs is one of the central points in Economy that has attracted the attention of many researchers. Since Pareto in 1897[1] found a cumulative power law distribution for several countries, many other studies have been reported, frequently with controversial results. Thus, while Pareto´s law has recently been observed for the distribution of individual income in Japan for fiscal year 1998 [2], studies of the American family income for 1947, 1973 and 1984 [3] and 1996 [4] clearly shows a maximum in the distribution that resembles the Maxwell-Boltzmann speed distribution of a monatomic gas [5], or in general, the Boltzmann distribution of energy in complex molecules [6]. It has also been proposed that these data follow an exponential function [4,7]. The USA personal income for the year 1935-36 was analysed by Montroll and Shlesinger who found that the top 1% follow a power low while the rest follows a log-normal distribution [8].

In a series of papers, Solomon et al showed that a stochastic dynamical model [9], based on the Lotka-Volterra system lead to power-law distributions [10 - 13].

Recently, an ideal gas model that takes into account the saving propensity predicted that the distribution of money should follow an asymmetric Gibbs-Boltzmann function [14].

The available information seems therefore to be rather contradictory while the theoretical basis of the nature of the distribution function and its origin, that should be used for its interpretation are not, at least, clear. Within this context, two pioneer studies were aimed to provide an insight on this subject using the principles and foundation of statistical physics. In one of them [15], a dynamical model of capital exchange was introduced which considered the transference of a specified amount of money between economic agents. Many different situations were analysed and the results obtained for the distribution of money in random multiplicative exchange, clearly presented the shape of a Gamma function.

The other study advanced further in the application of statistical mechanics to money distribution [16]. The authors proposed that the probability distribution of money must follow the exponential Boltzmann-Gibbs law characterized by an effective temperature equal to the average amount of money per economic agent.

Following the ideas of these two works, the present report propose to advance further into the development of a conceptual framework for the scientific analysis of income distribution. The basic idea arises from the evidence that the income distribution for different countries, when properly analysed, clearly follows a Gamma function that can be associated to the Boltzmann distribution of energy in polyatomic molecules, instead of the simple exponential of an ideal monatomic gas.

The other point addressed here, is the application of the basic concepts of molecular energy transfer processes to the rate of money transference and its instantaneous distribution [17]. A master rate equation is presented in Sec. 3 in line with the general formulation used in energy transfer studies [18] together with some representative computations to illustrate the behaviour of economic systems evolving to the equilibrium distribution.

In the analysis that follows the income data are assumed to be representative of the monies the agents posses, which is certainly an approximation.

## 2 The distribution of money

Let us consider a system composed of a large number of persons, which we shall call from here on economic agents. In this ensemble every agent can access to different states, each one characterized by an amount on money $M_i$ ranging from zero to a high, undetermined value. At a given instant all the agents are distributed among the different states, so that $n_1$ agents have an amount of money $M_1$, $n_2$ agents have $M_2$, and so on. The total number of agents is the total population $N$ of the ensemble:

$$N = n_1(t) + n_2(t) + .... = \Sigma n_i(t) \tag{1}$$

The first assumption is that $N$ is constant in time, so that there is neither decrease nor increment of population. In a real system this means that the newly incorporated agents plus immigration exactly compensate the number of deaths and migrations out of the ensemble. At any specified instant, $t$, the total amount of money in the system, $M$, is the sum of the money possessed by every agent:

$$M(t) = n_1(t)M_1 + n_2(t)M_2 + .... = \Sigma n_i(t)M_i \tag{2}$$

In real cases, the system is open and $M(t)$ is not constant but varies according to the flow on money from and into the country as a consequence of trade with other countries, the attitude of local and foreign investors and, in general, transactions with the rest of the world. Even though it is not strictly necessary, we will assume, for simplicity, that the external balance equals to zero and $M(t) = M$, which makes our system closed. In this system agents undergo pair wise interactions, freely or not, according to regulations, and the result of every interaction is the transference of money from one agent to another, so that as one of them increases the money possesses by an amount $\Delta M$, the other find it reduced by the same amount and the total money remains constant. The possible values of $\Delta M$ go from zero, when the transaction fails, up to a large, undefined value, less than $M$. We shall call $P_{ji}$ to the probability that an interaction could result in a transference $\Delta M$ so that and agent changes his money from $M_i$ to $M_j$, so that

$$M_i = M_j + \Delta M \tag{3}$$

In other words, $P_{ji}$ is the probability that an agent changes from state $i$ to $j$, as a consequence of a transaction. The rate of money transference $R$, is then the product of the rate of interaction between the agents, $\omega$, times the probability $P_{ij}$ that the interaction results in the transference of money $\Delta M$ from $j$ to $i$.

Therefore, there are in this system two different aspects that must be considered:

a) The rate at which money is transferred between the agents of the ensemble and

b) The distribution of money at any instant and when the system eventually reaches equilibrium.

In this section we will concentrate in the equilibrium distribution while the time evolution of the population will be analysed in Sec. 3, in connection with the rate of money transference.

Statistical analysis shows that the characteristic of any assembly of a large number of molecules $N$ is the emergence of a predominant configuration whose relevance increases sharply with $N$. This predominant configuration can be fully defined by a functional relation between the energy and the population of each level, which is an expression for the most probable number of molecules in level $i$, known as the Boltzmann distribution law of molecules:

$$n_i = A g_i \exp(-E_i / \beta) \qquad (4)$$

where $n_i$ is a probability density function, $A$ is a normalization factor, $g_i$ is the degeneracy or statistical weight of the energy level $E_i$ and $\beta$ is a constant with value $kT$, where $k$ is the Bolztmann constant and $T$ the absolute temperature. The constant $\beta$ is related to the average energy of the system, $<E>$. Thus, for an ideal monatomic gas, the average kinetic energy for each degree of freedom is $<E> = kT/2 = \beta/2$.

Based on the evidence provided by the available statistical information of income and the work of other authors, let us introduce as a second postulate that, in principle, the same statistical reasoning that leads to the Boltzmann distribution of energy in molecules applies to the distribution of money in an ensemble of economic agents and so, the equilibrium configuration is described by

$$n_i = A g_i \exp(-M_i / \beta) \qquad (5)$$

Here $M_i$ is the amount on money of state $i$ and $\beta$ should be related to the average money owned by the agents of the ensemble, $<M>$, to be consistent with the statistical mechanics of molecules.

The application of a Boltzmann distribution of energy to income distribution has been suggested previously, considering a non-degenerated situation, that is, $g_i = 1$, which corresponds to a monotonically decrease of $n_i$ with $M_i$ [4]. There are however indications that $g_i$ should be different from unit. In fact, a careful analysis of income distribution for various countries show a maximum, which clearly indicates a dependence of $g_i$ on $M_i$. A few representative cases are presented in Fig. 1. The data were obtained from the statistical information provided by the revenue services of Japan [2], the United Kingdom [19] and New Zealand [20]. Since the available data are given in different-sized bins they were reprocessed to obtain the normalized count, which is defined as the count in the class divided by the number of observations times the class width [21]. This is the appropriate normalization since we are using the histograms to model a probability density function, as required by Eq. (5). The money scale is given in thousands of New Zealand Dollars and in order to show the distributions from the three countries in the same graph, the data corresponding to the United Kingdom were divided by 2 and those from Japan by 2,000.

For the distribution of energy in ideal polyatomic molecules, the degeneracy of states can be calculated by closed expressions, such as for rotations under the rigid rotor approximation or by direct count or approximated methods for the normal modes of vibration [22]. In both of these examples, as in the general case, $g_i$ increases with energy.

At variance with molecules, $g_i$ cannot be computed *a priori* for economic systems. Due to this limitation, we used it as an adjustable parameter, assuming that $g_i$ increases with $M_i$ according to a potential law, as given by the Gamma function:

$$n_i = A(M_i)^{\alpha-1} \exp(-M_i/\beta) \qquad (6)$$

where $A$ is a normalization factor and $g_i = (M_i)^{\alpha-1}$.

In addition to the above expression for $g_i$, several other functional dependences were tried, but as they were more complex and did not produce any significant improvement on the fit, they are not presented here.

The results obtained for the fit to the data shown in Fig. 1 are given in Table I.

It is illustrative to anticipate here that an analysis of the income distribution for these countries in different years yields values of the parameters $\alpha$ and $\beta$ very close to those given in Table I, which indicates rather stable economies, close to equilibrium, or at least in stationary states, as expected if Eq. (5) is obeyed.

One of the advantages of the Gamma function is that its properties are well known. The average money value is simply the product $\alpha\beta$ and the width of the distribution, as measured by the variance is $\sigma^2 = \alpha\beta^2$. Computations of probability densities for different values of $\alpha$ and $\beta$ at a constant average of money of 100 arbitrary units are presented in Fig. 2. The results show that as $\alpha$ is increased and $\beta$ is consequently decreased as to keep the product invariant, the shape of the curve changes from that corresponding to a simple exponential decay to a function with a maximum. As $\alpha$ increases this maximum shifts to higher values of money while the function is more symmetric and narrow, that is, the ensemble becomes richer and more egalitarian.

## 3 The rate of money transference

In this section a deterministic approach to the rate of money transfer is presented. In order to analyse the rate at which money is transferred in and out of a specified money level $i$ it is necessary to consider the various processes by which agents enter or leave it. Following the standard procedure of energy transfer studies [22], these are:

1) Interaction between the agents to transfer money into level $i$.
2) Interaction between the agents to transfer money of the ensemble out of level $i$.
3) External input to the ensemble.
4) Output to the exterior of the ensemble.
5) Transfer out of $i$ by application of taxes.
6) Transfer into $i$ by redistribution of money collected from taxes.
7) Any other input and output mechanism.

Interaction between the agents results in the transference of money into level $i$ (Process 1) at a rate given by

$$R_{in} = \omega(P_{i1}n_1 + P_{i2}n_2 + ...) = \omega \sum_j P_{ij} n_j \qquad (7)$$

while the rate of transference out of level $i$ into any other level, process 2, is

$$R_{out} = \omega(P_{1i}n_i + P_{2i}n_i + ...) = \omega n_i \sum_j P_{ji} \qquad (8)$$

Since all the transactions from level $i$ must end at some other level, it is necessary that

$$\sum_j P_{ji} = 1 \qquad (9)$$

which makes the output rate equal to $\omega n_i$.

The rate of external money input into level $i$ of the ensemble can be expressed as $R_{ext\ in} f_i$, were $R_{ext\ in}$ is the total rate and $f_i$ is the fraction that enters into the $i$th level and the rate of transference of money outside the ensemble from level $i$ is $R_i$.

Similarly, in order to consider the effect of application of taxes two terms should be included. The rate at which the government collects money from the population in level $i$th is given by $R_{out, tax} = k_i n_i$, where $k_i$ depends of the way the agents $n_i$ are taxed. This money is then returned to the agents in various ways, such as salaries of the public employees, social security, etc. at a rate $R_{in, tax} h_i$ where $h_i$ is the fraction that goes into $i$.

The master rate equation for the system is then

$$dn_i / dt = R_{ext,in} f_i - R_i - \omega n_i + \omega \sum_j P_{ij} n_j - k_i n_i + R_{in,tax} g_i \qquad (10)$$

If the system is closed or, equivalently, the first two terms on the right are approximately equal and neglecting now, for simplicity, the effects of taxes, the above equation reduces to

$$dn_i / dt =_i -\omega n_i + \omega \sum_j P_{ij} n_j \qquad (11)$$

This equation gives the time evolution of the population of level $i$, in absence of any process except transference of money between the agents, to the final stationary value which is reached when $dn_i/dt=0$.

In order to solve the above master equation it is necessary to know the values of the elements $P_{ij}$. The problem is then exactly the same than that of energy transfer between gas molecules. As in that case, the values of $P_{ij}$ are not *a priori* known and the usual approach is to use various models in order to encompass a wide spectrum of possibilities. In the simplest model, which is usually called the stepladder model, interactions are assumed to add or remove a single amount of money, $\Delta M$, to every level, *i.e.*, all the transactions have the same value, which is the average amount of money exchanged in the ensemble. In this models the transition probability for the loss of money, from level $i$ to level $j$ is 1 if $M_j - M_i = \Delta M$ and zero otherwise. The elements $P_{ji}$ are subject to the condition of self-consistency, so that $P_{i+\Delta M,i} + P_{i-\Delta m,i} = 1$ as required by Eq. (9). $\Delta M$ is then the average value of money lost per transaction, $<\Delta M>_l$.

Another widely used model in energy transfer processes that seems worth using now for comparison is the exponential model. In this model the probability of removing the amount of money $\Delta M$ is proportional to $\exp(-\Delta M/<\Delta M>_l)$.

The usual practice in energy transfer studies is to treat $<\Delta M>_l$ as a purely empirical parameter selected to fit the experimental results.

The values of the transition probabilities that result in a gain of energy from the $i$th level are computed imposing the condition of detailed balance for the system in equilibrium, which is our third postulate. This condition relies on the principle of microscopic reversibility, which should always hold since it is based in the invariance of time reversal. In Physics, detailed balancing results from the existence of a Boltzmann distribution of energy, and consequently, for the problem considered now, we assume that it is applicable to the distribution of money. Then

$$P_{ji}/P_{ij} = \left(n_j/n_i\right)_{eqm} = \left(g_j/g_m\right)\exp\left[-\left(M_j - M_i\right)/\langle M \rangle\right] \qquad (12)$$

There is a main difference between energy transfer studies and the transference of money. In energy transfer studies the excited species are immersed in a bath of cold and inert molecules and relaxation takes place to reach the equilibrium distribution at the temperature of the bath gas. On the contrary, in a closed ensemble of agents who interchange money in pair wise transactions, the system evolves from an arbitrary initial composition to the final Boltzmann equilibrium at constant total money, so that the average amount of money transferred is zero, that is

$$dM/dt = \omega\langle \Delta M \rangle = 0 \qquad (13)$$

and the system evolves changing the values of the populations $n_i$, at constant $M$.

The results of two representative calculations are shown in Fig. 3, or the stepladder and the exponential models, with the same value of $<\Delta M>_l = 1$ a.u., starting from an initial narrow Gaussian distribution. The initial average money of the ensemble is 150 a.u. The calculations show that even though the intermediate states of the evolution are different, in both cases the system reaches the same final state, which is the Boltzmann distribution. The intermediate states can be satisfactorily fitted to Gaussian functions up to times close to the equilibrium. Similar results were obtained when the initial state is represented by a strictly singled valued Dirac function or any other probability density.

The immediate conclusion is that the equilibrium distribution of money in a closed society is independent of the way the process take place or, in other words, it cannot be affected by external factors, as long as the evolution of the system is conditioned by the principle of detailed balance. Then, irrespective of the constrains imposed by the initial distribution, the model for $P_{ij}$ and the corresponding parameterisation, the Boltzmann distribution of money is always reached, as the calculations presented here show.

The independence of the evolution of money on the shape of the initial distribution agrees with the results of Montroll and Shuler [23] who firsts demonstrated that the evolution of energy for an ensemble of diatomic molecules is independent of the initial energy distribution. That result was then extended to the more general case of polyatomic molecules [24]. However, it should be noted, as mentioned before, that those works dealt with the relaxation of excited species interacting with a reservoir of cold molecules, which represents a substantial difference with the present calculations.

It must also be pointed out that the similarity of money and energy transfer processes as analysed in this work, makes worth a further comparative study on the evolution of the moments of the distribution [25].

## 4 Conclusions

This study shows that the Boltzmann distribution of energy for polyatomic molecules applies well to the distribution of money, when the statistical information is presented as probability density distributions. However, the available data is provided by revenue services or statistical offices of different countries as individual income distributions and it is therefore necessary to assume that the monies actually possessed by the agents are proportional to the income.

Another required suppositions refer to the distribution of money and the rate of money transference between the agents. The fundamental postulate is that money distribution is merely statistical and, as for molecules, the equilibrium distribution corresponds to the predominant configuration of the system, whose quantitative expression is the Bolztmann distribution law of molecules (Eq. (4)). In addition, microscopic reversibility is also a imposed on the system (Eq. (12)). As a consequence, even tough the probability densities at intermediate times during the evolution of the system are determined by the specific model for the transition probabilities, $P_{ji}$, the final state will be given by Eq.(5). That is, the final state is independent of the initial distribution and of the evolution pathway, as model calculations show.

The author thanks CONICET, SECYT , ACC and FONCYT for financial support.

**References**

[1] V. Pareto, *Le Cours d'Économie Politique* ( Macmillan, London, 1897).

[2] H. Aoyama, Y. Nagahara, M. P. Okazaki, W. Souma, H. Takayasu, M. Takayasu, Fractals **8**, 293 (2000)

[3] F. Levy, Science **236**, 923 (1987)

[4] A. A. Dragulescu , V. M. Yakovenko, Eur. Phys. J. B **20**, 585 (2001)

[5] L. K. Nash, Elements of Statistical Thermodynamics (Addison-Welsey, 1972)

[6] R.C. Tolman, *Foundations of Statistical Mechanics* (Oxford University Press, 1938); N. Davidson, *Statistical Mechanics*, McGraw-Hill, New York, 1962)

[7] A. A. Dragulescu, V. M. Yakovenko, Phys. A **299**, 213 (2001)

[8] F. W. Montrol and M. F. Shlesinger, J. Stat. Phys. **32**, 209 (1983)

[9] O. Malcai, O. Biham, P. Richmond, S. Solomon, Phys. Rev. E **66**, 31102 (2002)

[10] M. Levy, S. Solomon, Physica A **242**, 90 (1997)

[11] S. Solomon, M. Levy, Int. J. Mod. Phys. C **7**, 745 (1996)

[12] *Elements of physical Biology*, edited by A. J. Lokta (Williams and Wilkins, Baltimore, 1925)

[13] V. Volterra, Nature **118**, 558 (1926)

[14] B. K. Chakrabarti , A. Chartterjee, arXiv:cond-mat/0302147.

[15] S. Ispolatov, P. L. Krapivsky , S. Redner, Eur. Phys. Jour. B **2**, 267 (1998)

[16] A. A. Dragulescu, V. M. Yakovenko, Eur. Phys. Jour. B **17**, 723 (2000)

[17] I. Oref, D. C. Tardy, Chem. Rev. **90**, 1407 (1990)

[18] C. A. Rinaldi, J. C. Ferrero, M. A. Vázquez, M. L. Azcárate, E. J. Quel, J. Phys. Chem. **100**, 9745 (1996); E. A. Coronado, J. C. Ferrero, J. Phys. Chem. **101**, 9603 (1997)

[19] Inland Revenue: Income distribution. http://www.inlandrevenue.gov.uk /stats/income_distribution/pi_t03_1.thm

[20] New Zealand Income Survey. http://www.stats.govt.nz

[21] Engineering Statistics Handbook. http://www.itl.nist.gov/div898/handbook

[22] P. J. Robinson, K.A. Holbrook, *Unimolecular Reactions* (Wiley, London, 1972)

[23] E. W. Montroll, E. K. Shuler, J. Chem. Phys. **26**, 454 (1957)

[24] E. A. Coronado, J. C. Ferrero, Chem. Phys. Lett. **227**, 164 (1994)

[25] E. A. Coronado and J. C. Ferrero, Chem. Phys. Lett. **257**, 674 (1966)

**Figure captions**

Figure 1. Income distribution for Japan, New Zealand and the UK. The income values for Japan has been scaled by a factor 500 and those for the UK by a factor of 2. The lines represent the best fit to the Gamma function, as indicated in the text.

Figure 2. Evolution of the probability density distribution calculated with different models of transaction probability, with $<\Delta M>=1$ and a narrow initial Gaussian distribution centred at M = 150. The values are in arbitrary units. The lines shown correspond to the same time for both calculations, from top to bottom. The left figure show the results obtained using an exponential model while the right figure corresponds to a step-ladder model. The broad line represents the Bolztmann equilibrium distribution.

Figure 3. Probability density distribution calculated using the Gamma function with different values of $\alpha$ and $\beta$, but with the same average value of M: ( — ) $\alpha=1$, $\beta=100$; ( ---- ) $\alpha=1.5$, $\beta=66.7$; (·····) $\alpha=2$, $\beta=50$; (− · −) $\alpha=3$, $\beta=33.3$; (· · −) $\alpha=10$, $\beta=10$.

Table 1. Values of α and β corresponding to Eq. (6) obtained from the fits of the income data for New Zealand (1998), UK (1998-99) and Japan (1996)

|  | α | β |
|---|---|---|
| New Zealand | 1.60 | 13.50 kNZD |
| United Kingdom | 1.98 | 8.03 k£ |
| Japan | 2.01 | 2.67 M¥ |

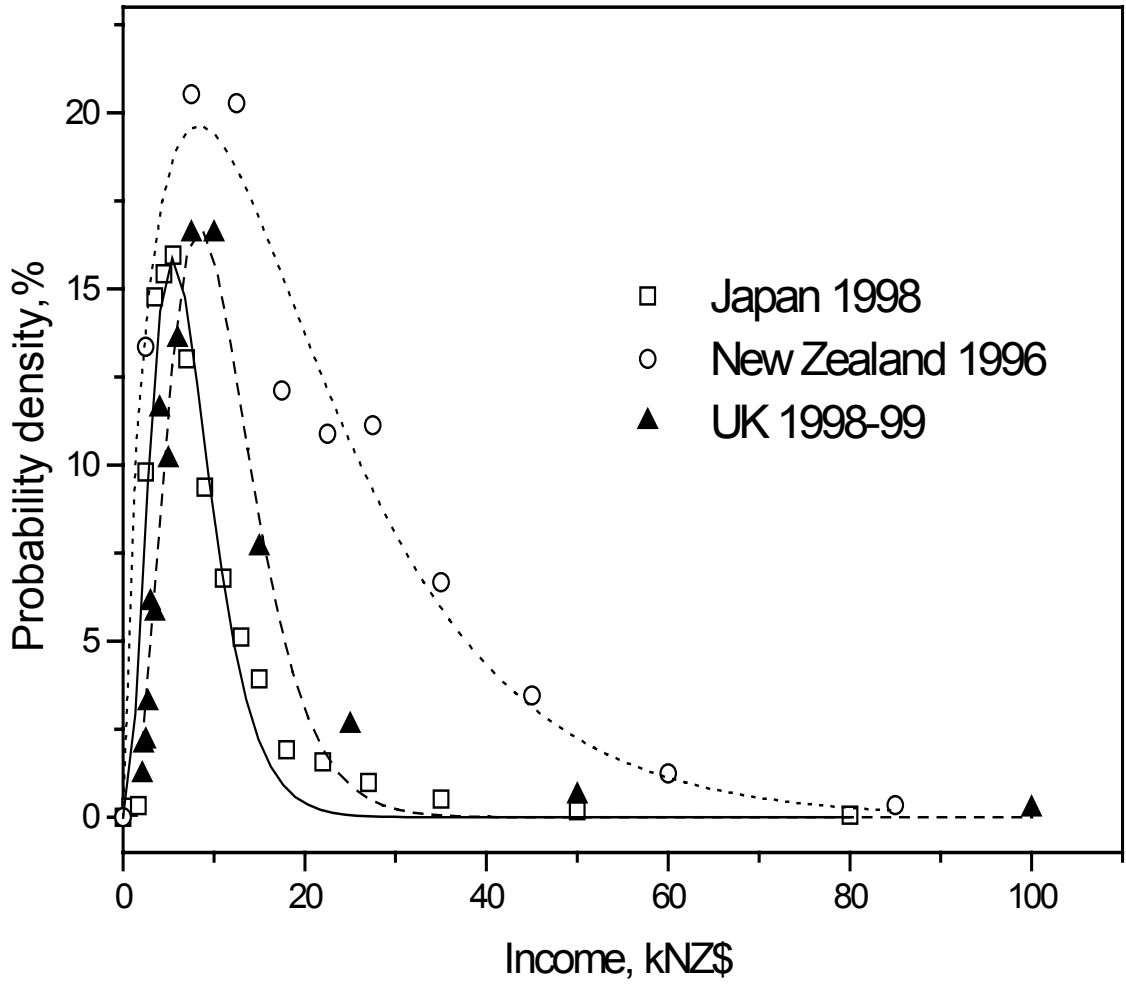

Figure 1

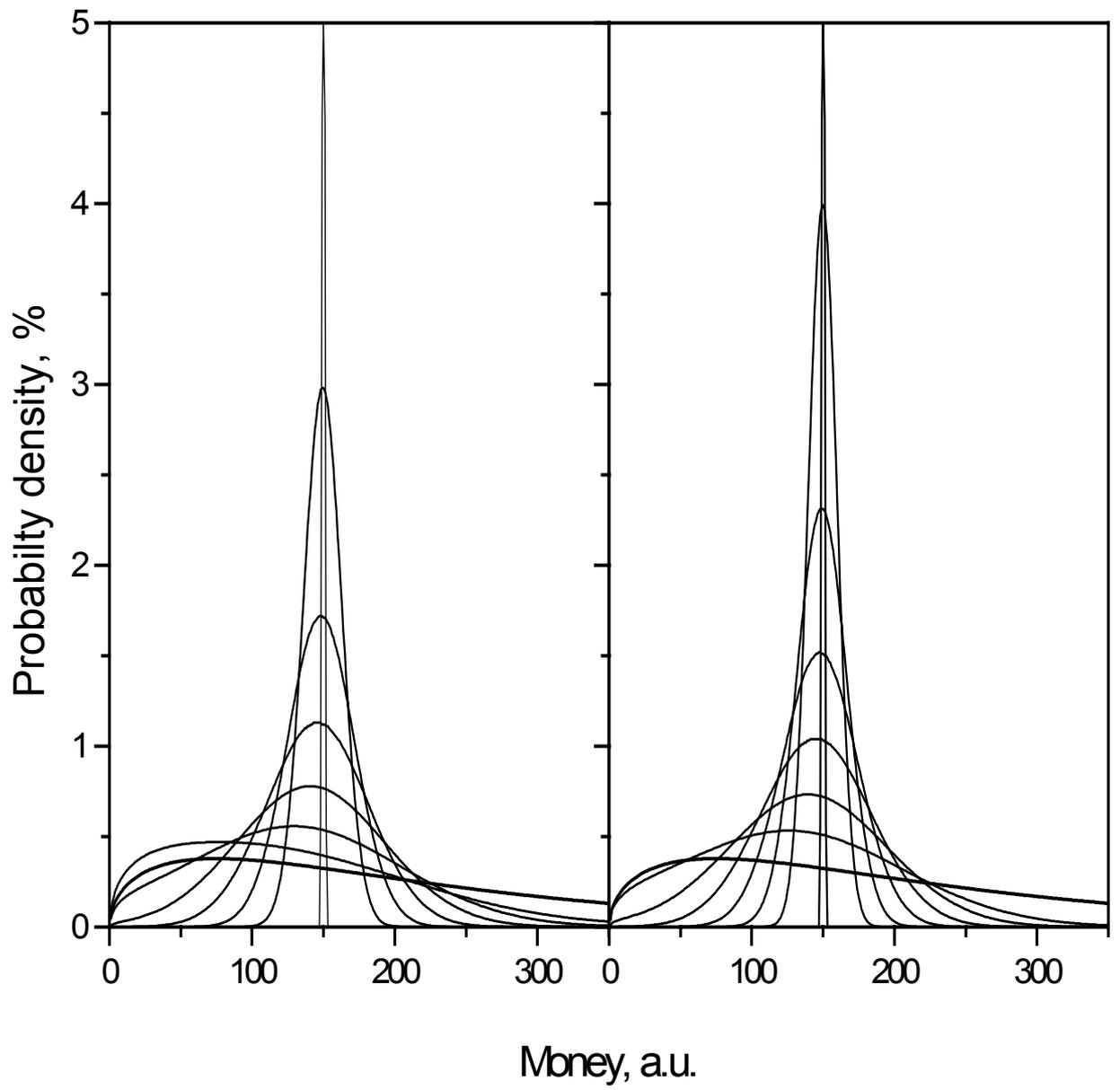

Figure 2

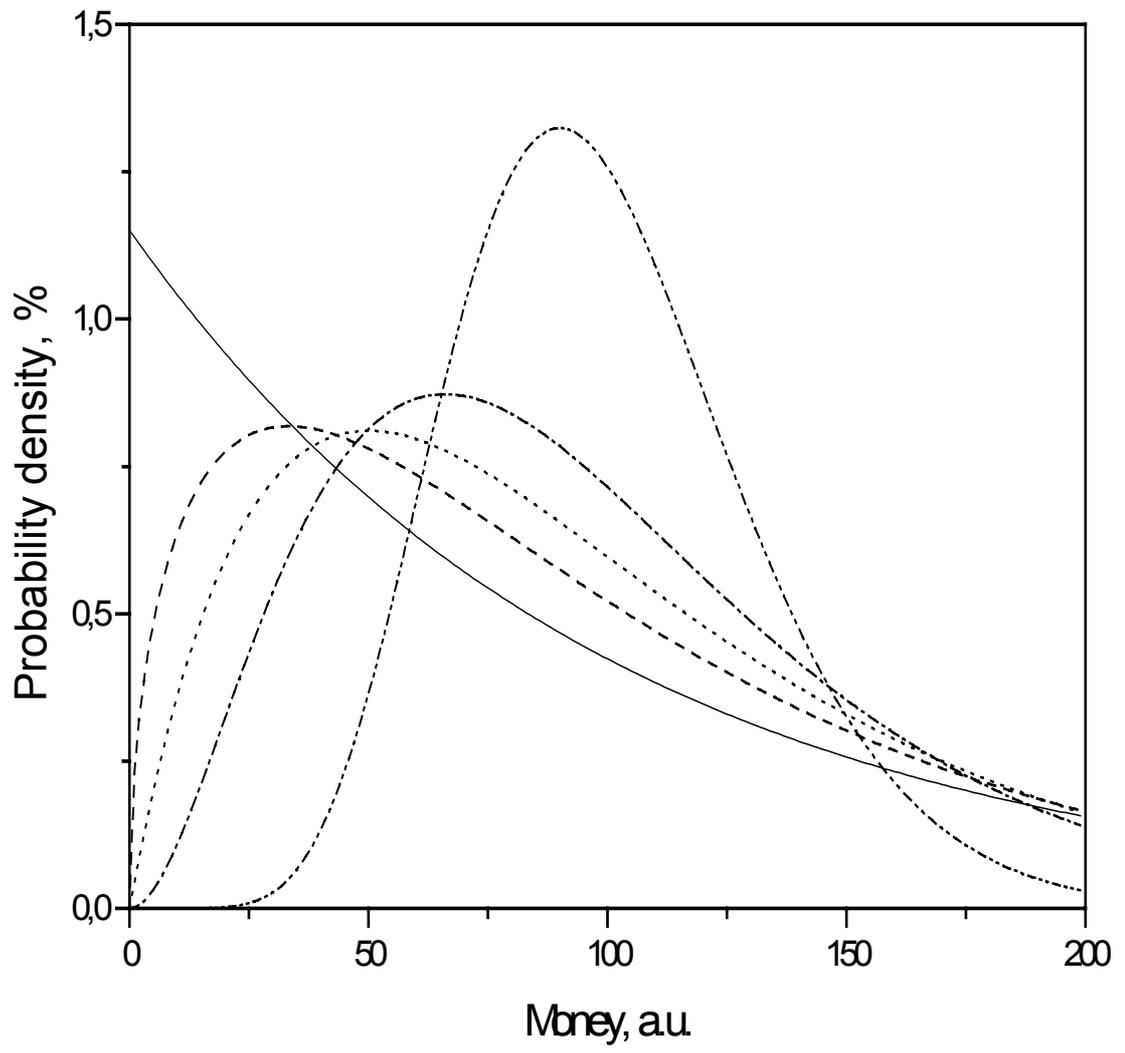

Figure 3